\pdfoutput=1
\documentclass[final,5p,times,twocolumn]{elsarticle}
\usepackage{lineno}
\usepackage[bookmarks=true,pdfpagelabels,pdfborder={0 0 0},breaklinks=true,colorlinks=true]{hyperref}
\usepackage{amsfonts}
\usepackage[centertags]{amsmath}
\usepackage{amssymb,amscd,amsbsy}
\usepackage{eurosym}
\usepackage{upgreek}
\usepackage{paralist}
\usepackage{array}
\usepackage{ulem}
\usepackage{lastpage}
\usepackage[font=small]{caption}
\usepackage{calc}
\journal{Nuclear Instruments \& Methods in Physics Research, Section A}
\graphicspath{{./}{./pictures/}}
\begin{document}
\begin{frontmatter}
\title{Cherenkov light imaging in astroparticle physics}

\author{U.\,F.~Katz}
\address{Erlangen Centre for Astroparticle Physics, 
         Friedrich-Alexander University Erlangen-N\"urnberg,
         Erwin-Rommel-Str.~1, 91058 Erlangen, Germany}
\ead{katz@physik.uni-erlangen.de}

\begin{abstract}
Cherenkov light induced by fast charged particles in transparent dielectric
media such as air or water is exploited by a variety of experimental techniques
to detect and measure extraterrestrial particles impinging on Earth. A selection
of detection principles is discussed and corresponding experiments are presented
together with breakthrough-results they achieved. Some future developments are
highlighted.
\end{abstract}

\begin{keyword}
Astroparticle physics\sep Cherenkov detectors\sep neutrino telescopes\sep
gamma-ray telescopes\sep cosmic-ray detectors
\end{keyword}
\end{frontmatter}
%
%
\section{Introduction}
\label{sec-int}

In 2018, we commemorate the 60$^\text{th}$ anniversary of the award of the Nobel
Prize to Pavel Alexeyewich Cherenkov, Ilya Mikhailovich Frank and Igor
Yevgenyevich Tamm for {\it the discovery and the interpretation of the Cherenkov
effect} \cite{Cherenkov1934,FrankTamm1937}. The impact of this discovery on
astroparticle physics is enormous and persistent. Cherenkov detection techniques
were instumental for the dicovery of neutrino oscillations; the detection of
high-energy cosmic neutrinos; the establishment of ground-based gamma-ray
astronomy; and important for the progress in cosmic-ray physics.

The characteristics of Cherenkov radiation are governed by its emission angle with
respect to the particle's direction of flight, $\cos\theta_\text{C}=1/(n\beta)$
($n$ and $\beta=v/c$ being the refractive index and the particle velocity,
respectively) and by its intensity, given by the Frank-Tamm formula
\begin{equation*}
\begin{split}
\frac{\text{d}N_\gamma}{\text{d}x}&=
2\pi\alpha\left(1-\frac1{\beta^2n^2}\right)\cdot
\left(\frac1{\lambda_\text{min}}-\frac1{\lambda_\text{max}}\right)\\
&\stackrel{\beta=1}{\approx}
\begin{cases}
3\times10^4/\text{m}&\text{in water/ice}\\
15/\text{m}           &\text{in air (8\,km height)}
\end{cases}\\
&\text{for}\quad\lambda_\text{min}=300\,\text{nm}\le\lambda\le
                \lambda_\text{max}=600\,\text{nm\,.}
\end{split}
\end{equation*}
Here, $\lambda$ is the emitted wavelength and the $\lambda$ range indicated
roughly corresponds to the sensitivity range of typical light sensors. The
geometry of Cherenkov emission allows for reconstructing the particle
trajectory, provided sufficiently many Cherenkov photons are measured with good
spatial and time resolution, and they can be separated from background light. 
The recorded Cherenkov intensity furthermore can serve as a proxy for the
particle energy. Usually, the detectors need to be shielded from ambient light
and employ photo-sensors that are sensitive to single photons with nanosecond
time resolution. Photomultiplier tubes (PMTs) and, more recently, silicon
photomultipliers (SiPMs) \cite{Barbato,Vinogradov} are the standard sensor
types. They are provided by specialised companies who cooperate with the
experiments in developing and optimising sensors according to the respective
specific needs (see e.g.\ \cite{Ziembicki,Sidorenkov}).

In the following, the detection principles of different types of Cherenkov
experiments in astroparticle physics are presented together with selected
technical details and outstanding results.
%
%
\section{Ground-based gamma-ray detectors}
\label{sec-gam}

While the atmosphere is transparent to electromagnetic radiation in the radio
and optical regimes, it is not for X-rays and gamma-rays. Gamma-rays below
20\,GeV are only accessible to satellite experiments. At significantly higher
energies, satellite instruments rapidly lose sensitivity to the steeply
decreasing gamma-ray flux due to their limited collection area, and ground-based
observations take over (see \cite{Funk2015} for more details). Imaging air
Cherenkov telescopes (IACTs) require clear, preferentially moon-less nights and
sites with negligible light pollution and an elevation of typically 2\,km. They
are pointing instruments with a field of view of a few degrees in diameter. 
Alternatively, timing arrays at higher altitude can directly measure the
gamma-induced particle cascade. They cover a significant fraction of the sky,
albeit with a higher energy threshold than IACTs and inferior sensitivity at
energies below about 50\,TeV.
%
%
\subsection{Detection principle of Cherenkov telescopes}
\label{sec-gam-det}

Upon entering the atmosphere, gamma-rays with GeV energies and above initiate
electromagnetic cascades, extending longitudinally over several kilometres and
having their maximum typically at a height of 10\,km above sea level. The
integrated track length of all $e^\pm$ in the cascade and therefore the overall
Cherenkov light yield is to a good approximation proportional to the initial
gamma-ray energy, $E_\gamma$. The Cherenkov angle is about
$\theta_\text{C}=1^\circ$ and increases with the air density, i.e.\ along the
cascade. Even though the basic processes in the cascade -- pair creation and
bremsstrahlung -- are very close to being collinear with the incoming gamma-ray,
multiple scattering of the $e^\pm$ widens the Cherenkov light pool, which is
concentrated within a radius of about 100--150\,m at ground level.

\begin{figure}[ht]
\begin{center}
\begin{minipage}{\columnwidth}
  \includegraphics[width=\columnwidth]{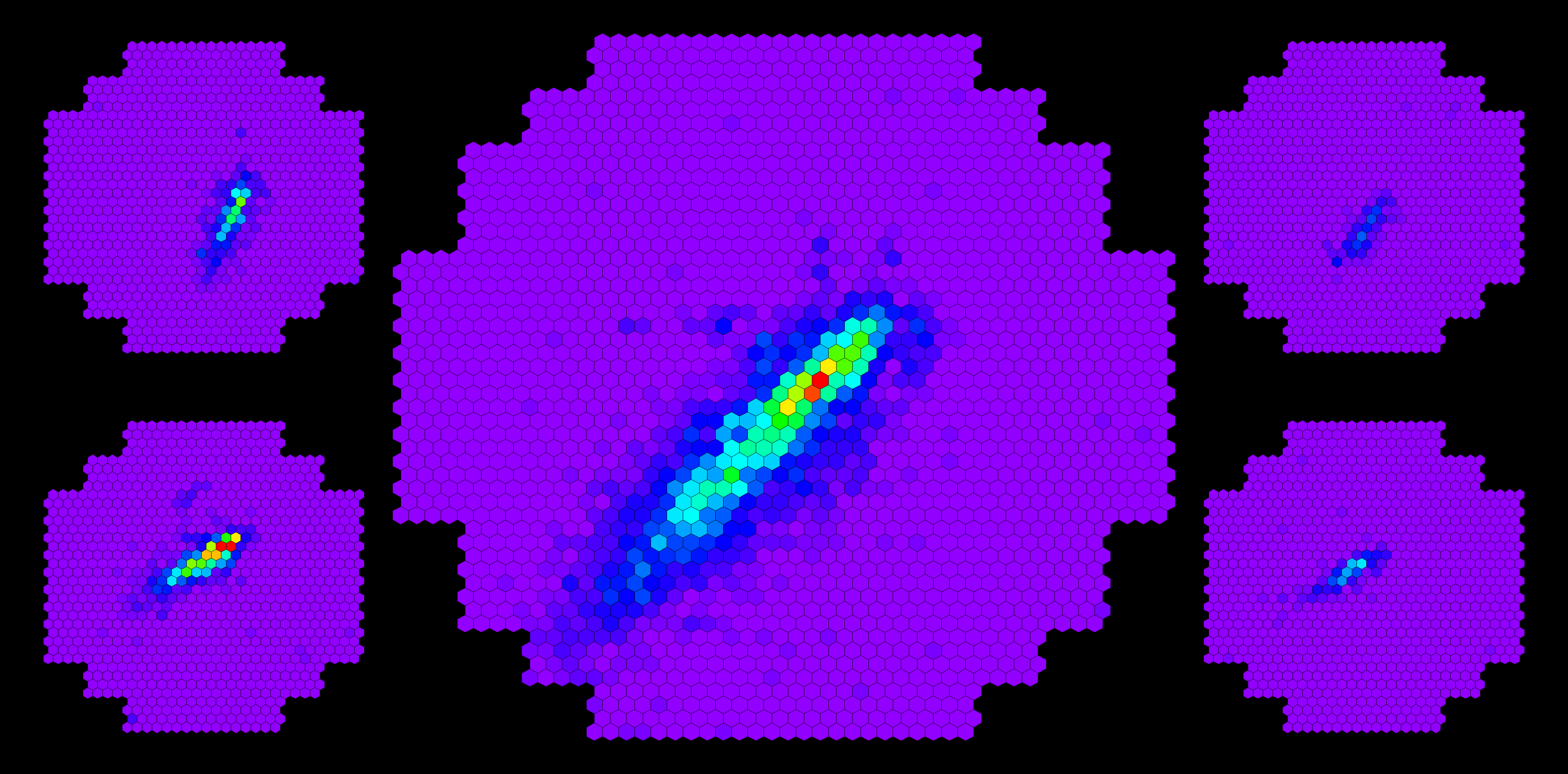}
\end{minipage}%
\caption{%
Camera pictures of a gamma-ray shower simultaneously observed by all five
H.E.S.S.\ telescopes. Each hexagonal pixel corresponds to one photomultiplier. 
The colour code indicates the measured light intensity. Picture provided by the
H.E.S.S.\ Collaboration.}
\label{fig-hess-cam}
\end{center}
\end{figure}

Since all relevant particles in the cascade propagate with a speed very close to
that of light, the Cherenkov radiation arrives at ground in a flash of only a
few nanoseconds duration and can thus be separated from the night-sky
background. It is observed with one or several IACTs that have a camera made of
photomultiplier or SiPM pixels in their focal plane. An important parameter,
governing e.g.\ the ability for large-area sky scans, is the field of view of
the camera (a few degrees). Each camera pixel corresponds to a certain solid
angle of the light arrival direction. A telescope pointed to a gamma-ray source
thus sees the start of the cascade (small $\theta_\text{C}$, little multiple
scattering) close to its centre, from where it propagates outward. An example of
a cascade observed by all five H.E.S.S.\ telescopes (see
Sect.~\ref{sec-gam-cur}) is shown in Fig.~\ref{fig-hess-cam}. The gamma-ray
direction and energy are reconstructed from the recorded light pattern and
intensity. Typical resolutions of $0.1^\circ$ and 20\%, respectively, are
achieved for stereoscopic observations by several cameras. Atmospheric cascades
induced by cosmic-ray protons or heavier nuclei are three orders of magnitude
more frequent than gamma-rays, but can be efficiently suppressed using the
topology of the camera pictures (hadronic showers are more fuzzy).

\begin{figure}[ht]
\begin{center}
\begin{minipage}{\columnwidth}
  \includegraphics[width=\columnwidth]{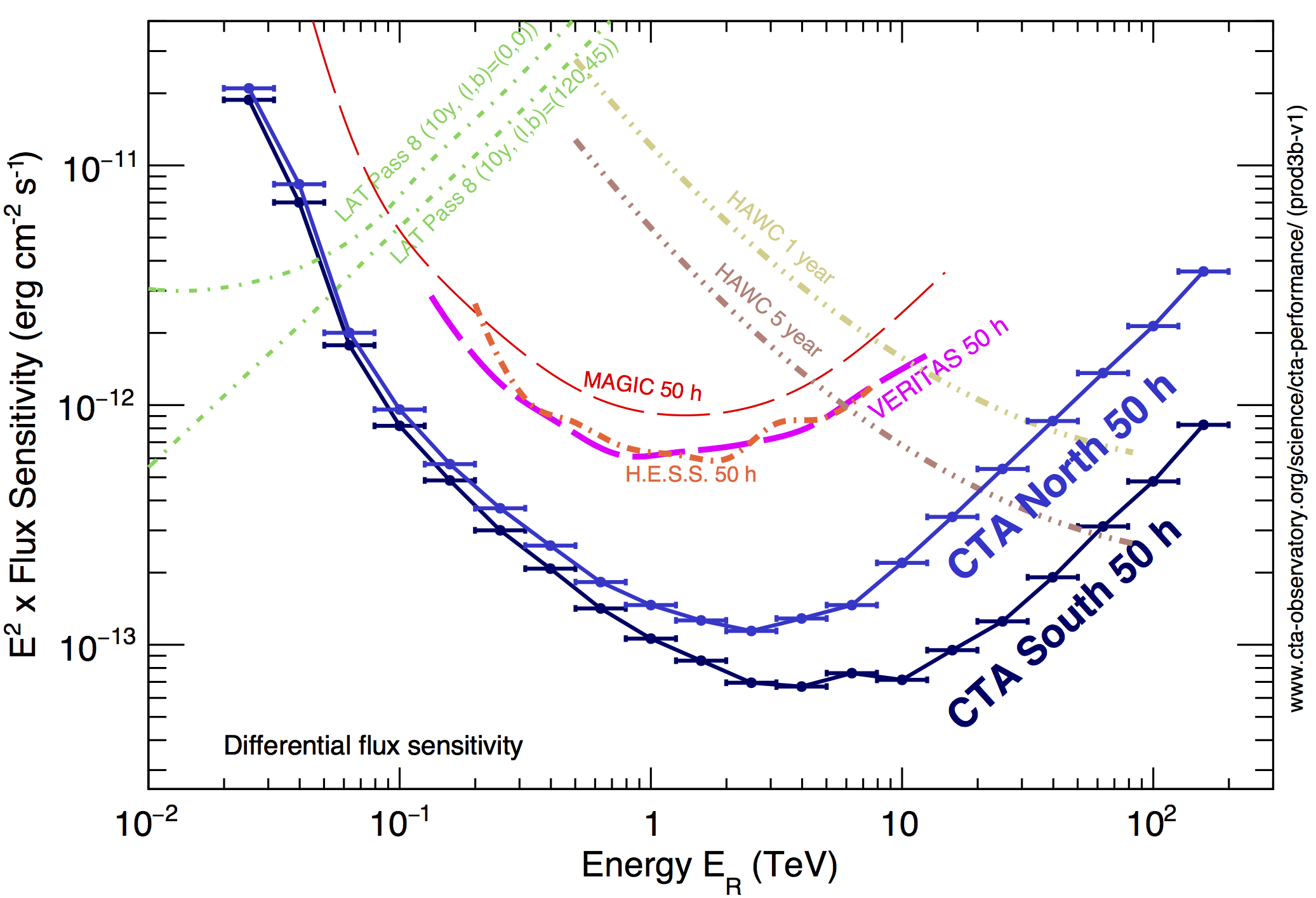}
\end{minipage}%
\caption{%
Differential flux sensitivity of the current (H.E.S.S., MAGIC, VERITAS) and the
future (CTA) ground-based IACTs. Also shown are the corresponding sensitivities
for the timing array HAWC. See Sections~\ref{sec-gam-cur}--\ref{sec-gam-tim} for
more details on these instruments. The green, dash-dotted lines indicate the
sensitivity of the satellite instrument Fermi-LAT for two different directions
of observation. See Sections~\ref{sec-gam-cur}--\ref{sec-gam-tim} for more
details on these instruments. Note that the vertical axis shows
$E_\gamma^2\cdot\text{flux}$. Picture provided by the CTA Collaboration.}
\label{fig-cta-sens}
\end{center}
\end{figure}

The observable $E_\gamma$ range is limited by the light intensity and the background
separation at low energies (some 10\,GeV for the largest telescopes in
operation) and by the overall collection area at high energies (about 100\,TeV
for a large array of smaller telescopes). The sensitivities of current and
future gamma-ray telescopes are compared in Fig.~\ref{fig-cta-sens} as functions
of~$E_\gamma$.

\begin{table}[ht]
\begin{center}
\caption{%
Main parameters of the currently operational IACTs and references to their web
pages. The second MAGIC telescope came in operation 2009, the large H.E.S.S.\
telescope in 2012.}
\renewcommand{\arraystretch}{1.26}
\begin{tabular}{lccc}
\hline
                &H.E.S.S.       &MAGIC          &Veritas\\
\hline
Site            &Namibia        &La Palma,      &Arizona, US\\[-1.mm]
                &               &Spain          &\\
\hline
Altitude a.s.l. &1800\,m        &2200\,m        &1270\,m\\
\hline
Operation       &2003--         &2004--         &2007--\\
\hline
Telescopes      &5              &2              &4\\
\hline
Dish diameter   &4$\times$12\,m,&2$\times$17\,m &4$\times$12\,m\\[-1.mm]
                &1$\times$28\,m &               &\\
\hline
Field of view   &5$^\circ$      &3.5$^\circ$    &3.5$^\circ$\\
\hline
Photo-sensors   &PMTs           &PMTs           &PMTs\\
\hline
Web page        &\cite{HESS}    &\cite{MAGIC}   &\cite{Veritas}\\
\hline
\end{tabular}
\label{tab-iacts}
\end{center}
\end{table}
%
%
\subsection{Current Cherenkov telescopes}
\label{sec-gam-cur}

\begin{figure*}[t]
\begin{center}
\begin{minipage}{\textwidth}
  \centering
  \includegraphics[width=0.8\textwidth]{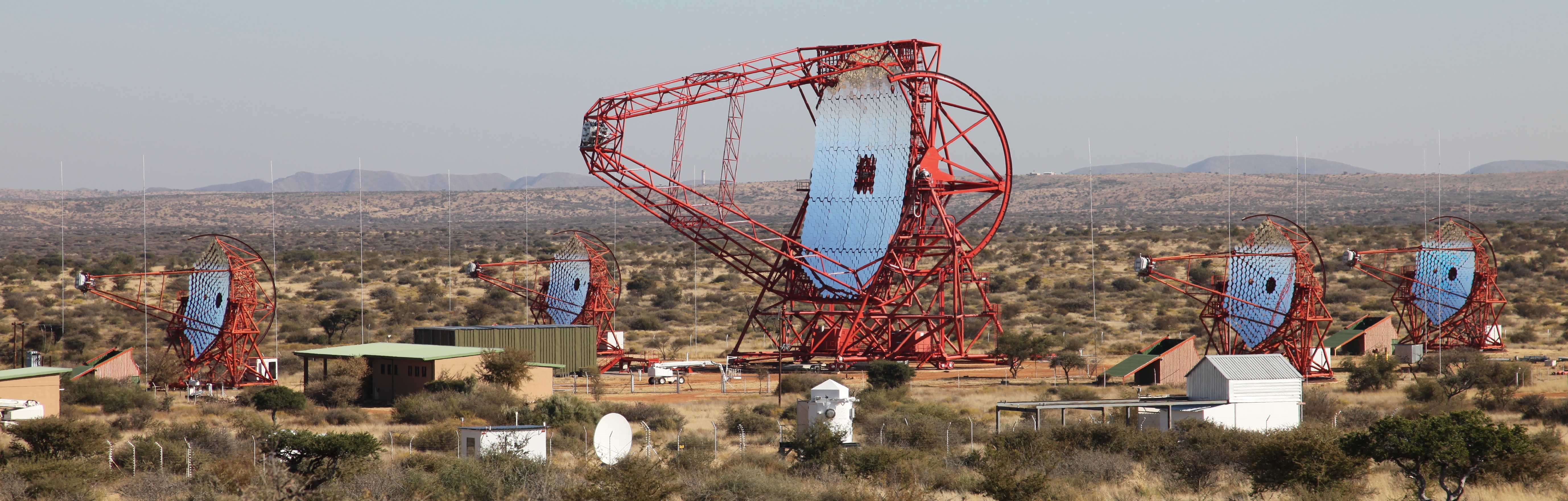}
\end{minipage}%
\caption{%
Photograph of the H.E.S.S.\ telescope system in Namibia. Picture provided by the
H.E.S.S.\ Collaboration.}
\label{fig-hess}
\end{center}
\end{figure*}

After the feasibility of ground-based gamma-ray astronomy had been demonstrated
in the late 1980's and 1990's by the Whipple \cite{Weekes1989} and HEGRA
\cite{Bradbury1997} instruments, a second generation of IACTs became operational
in the 2000's. The main properties of these telescopes are summarised in
Table~\ref{tab-iacts}; a photograph of the H.E.S.S.\ instrument is shown in
Fig.~\ref{fig-hess}. They have established gamma-ray astronomy as a major field
of observational astrophysics and provided a wealth of scientific information on
high-energy processes in the Universe. More than 200 gamma-ray sources in the
TeV regime have been detected and their spectra, light curves and -- for
Galactic sources -- morphologies investigated. Gamma-ray emission and thus
particle acceleration beyond TeV energies has been proven for a number of object
classes, amongst them shell-type supernova remnants, pulsar wind nebulae,
compact binaries and active galactic nuclei. See \cite{Funk2015} and references
therein for details.

In addition to the IACTs listed in Table~\ref{tab-iacts}, a first small
telescope with a camera equipped with SiPMs, FACT \cite{FACT}, has been
constructed and is operated since 2011 in La Palma for long-term monitoring
purposes. FACT has demonstrated that SiPMs can take very high rates, enabling
observations even in full-moon nights.
%
%
\subsection{Cherenkov Telescope Array}
\label{sec-gam-cta}

The H.E.S.S., MAGIC and Veritas collaborations have joined forces to construct
the next-generation IACT, the Cherenkov Telescope Array (CTA). CTA will be
installed in two sites, one in the Northern hemisphere on La Palma and the other
in the Southern hemisphere close to the ESO Paranal Observatory in Chile. CTA
will consist of telescopes of three sizes, LSTs (large size telescopes), MSTs
(medium) \cite{Barrio} and SSTs (small) \cite{Heller}, with a field of view of
$4.5^\circ$ (LST) and $8^\circ$ (MST and SST), respectively. At the Chile site,
they will be arranged in concentric groups of 4~LSTs in the middle, followed by
25~MSTs and 70 SSTs. On La Palma, there will be 4 LSTs and 15 MSTs, but no SSTs. 
This takes into account that on the Northern site predominantly extragalactic
observations will be made, where the gamma-ray flux beyond some 10\,TeV --
targeted by the SSTs -- is strongly reduced through absorption by the
extragalactic background light. The first LST was recently inaugurated in La
Palma. CTA is expected to start operation in 2022 with partial arrays and to be
completed in 2025. The CTA sensitivity (see Fig.~\ref{fig-cta-sens}) promises
major progress in gamma-ray astronomy and high-energy astrophysics once CTA will
take data.

The major scientific targets of CTA are cosmic particle acceleration, probing
extreme environments such as close to neutron stars and black holes, and
fundamental physics such as investigations into the nature of dark matter. CTA
will be operated as an observatory, with some key science projects reserved for
the CTA Collaboration. These include surveys of the Galactic Centre, the
Galactic Plane, the Magellanic Cloud, and extragalactic objects, as well as the
investigation of cosmic-ray PeVatrons, star-forming galaxies, active galactic
nuclei and clusters of galaxies. Also, the dark matter programme and the
exploitation of CTA data beyond gamma-rays are key science projects.
%
%
\subsection{Timing arrays}
\label{sec-gam-tim}

At altitudes exceeding about 4\,km above sea level, the particle cascades
induced by gamma-rays and cosmic rays at energies beyond 100\,GeV reach the ground
and can be observed with arrays of suitable detectors, e.g.\ water tanks in
which through-going charged particles generate Cherenkov light detected by PMTs. 
From measuring the arrival time of the shower front as a function of the
horizontal position, the direction of the incoming particle can be determined. 
The intensity of the shower and the size of its footprint on ground yield an
energy estimate. Similarly to IACTs, leptonic and hadronic showers are separated
using the different event topologies and muon content in the detector array.

\begin{figure}[ht]
\begin{center}
\begin{minipage}{\columnwidth}
  \includegraphics[width=\columnwidth]{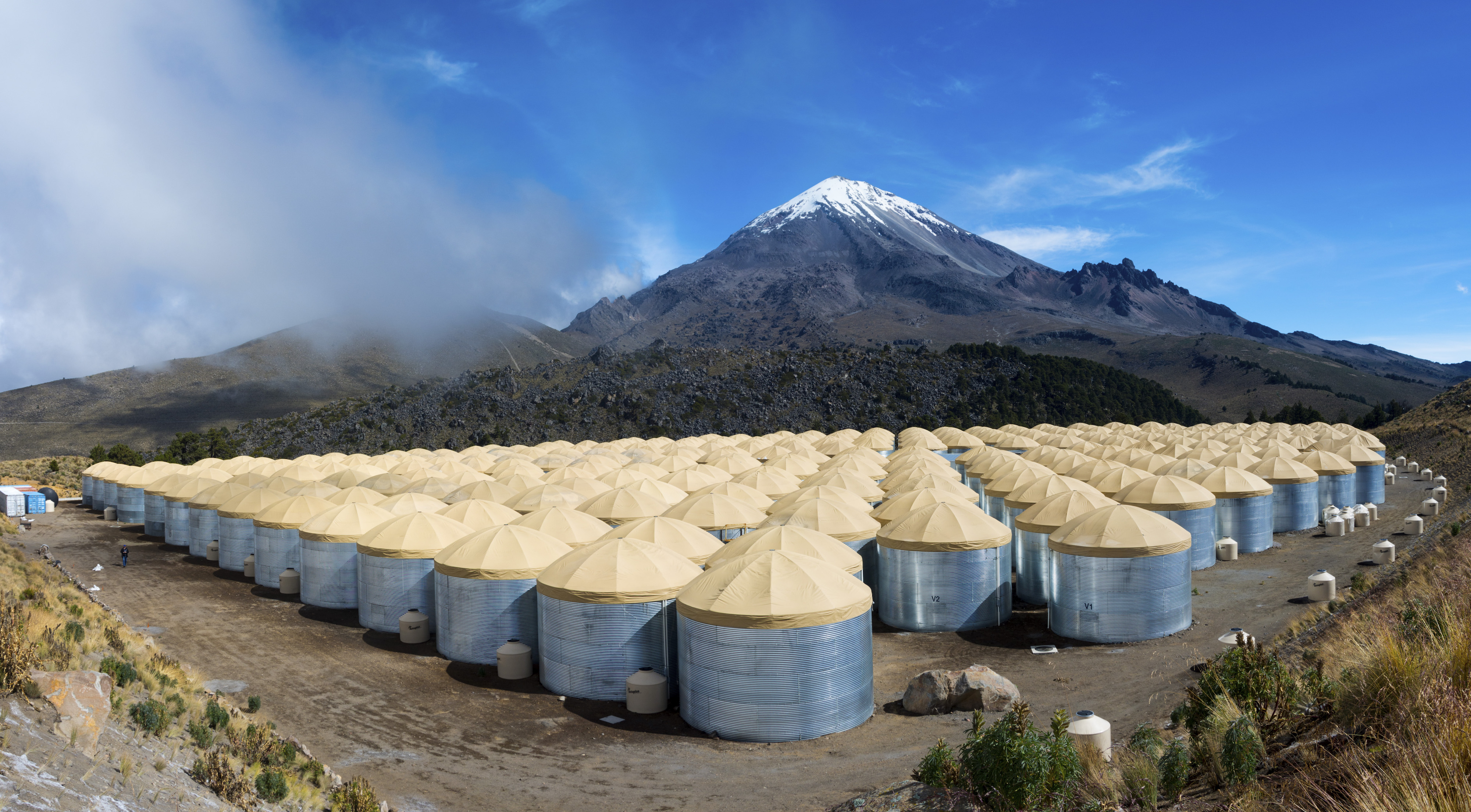}
\end{minipage}%
\caption{%
Photograph of the HAWC detector array in Mexico. Picture provided by the
HAWC Collaboration.}
\label{fig-hawc-photo}
\end{center}
\end{figure}

The currently most sensitive detector of this type is HAWC \cite{HAWC} near
Puebla in Mexico, at an altitude of 4100\,m. HAWC consists of 300 water tanks
covering $0.05\,\text{km}^2$, each filled with 200\,tons of ultra-pure water
observed by 4 PMTs. A photograph of HAWC is shown in Fig.~\ref{fig-hawc-photo}. 
HAWC reaches an angular resolution of about $0.1^\circ$ for gamma-ray energies
$E_\gamma\gtrsim10\,$TeV. Even though, for $E_\gamma\lesssim100\,$TeV, HAWC is
less sensitive than CTA for any given direction, it has the advantage of a large
field of view and a very high duty cycle. It is particularly well suited to
observe very high-energy gamma-ray emission from extended objects. As an
example, a recent measurement of the TeV gamma-ray flux from the vicinity of two
pulsars has strongly constrained the hypothesis that positrons from these/such
pulsars is responsible for the unexpectedly high flux of cosmic-ray positrons at
Earth \cite{HAWC2017}.
%
%
\section{Neutrino telescopes}
\label{sec-neu}

Due to their notoriously small interaction cross section, neutrinos are very
good, long-range astrophysical messengers; on the other hand, they are difficult
to detect. The basic principle of neutrino telescopes is to observe Cherenkov
light from charged secondary particles emerging from neutrino reactions and
passing a detector volume filled with a transparent dielectric medium and
observed by an arrangement of PMTs (due to their comparatively high noise rates,
SiPMs are not [yet] suited for neutrino telescopes). For the low-energy regime
(typically MeV--multi-GeV), detectors are installed in deep-underground caverns
and the PMTs cover a large percentage of the detector volume outer surface. For
high energies (some\,GeV--10\,PeV), naturally abundant volumes of water or ice
are instrumented with three-dimensional arrays of PMTs \cite{KatzSpiering2012}.
%
%
\subsection{Low-energy neutrino detectors}
\label{sec-neu-low}

The neutrino fluxes observed with theses detectors are those from the Sun
(4--20\,MeV), from supernovae (10--30\,MeV), atmospheric neutrinos generated in
cosmic-ray induced particle cascades (sub-GeV--TeV), and also beam neutrinos
from accelerators for long-baseline experiments (GeV). The energy ranges
indicate the typical observation windows of Cherenkov detectors. Note that the
dominant neutrino interactions observed in these cases are different: Solar
$\nu_e$ are detected via elastic $\nu_ee^-\to\nu_ee^-$ scattering, where the
final-state $e^-$ preserves the $\nu_e$ direction; supernova neutrinos are
mostly visible via $\overline{\nu_e}p\to e^+n$, with poor directional
information; atmospheric (anti)neutrinos produce high-energy $e^\pm$ or
$\mu^\pm$, the direction and particle type of which are measured in the
detector. The physics questions addressed through these measurements are
neutrino oscillations, the processes in the Sun and in supernovae, and searches
for relic neutrinos from unresolved supernovae and for possible sterile
neutrinos.

Two detectors have provided outstanding results: Super-Kamiokande (SK) in Japan
\cite{SK} and the Sudbury Neutrino Observatory (SNO) \cite{SNO} in Canada. 

\begin{figure}[t]
\begin{center}
\begin{minipage}{\columnwidth}
  \includegraphics[width=\columnwidth]{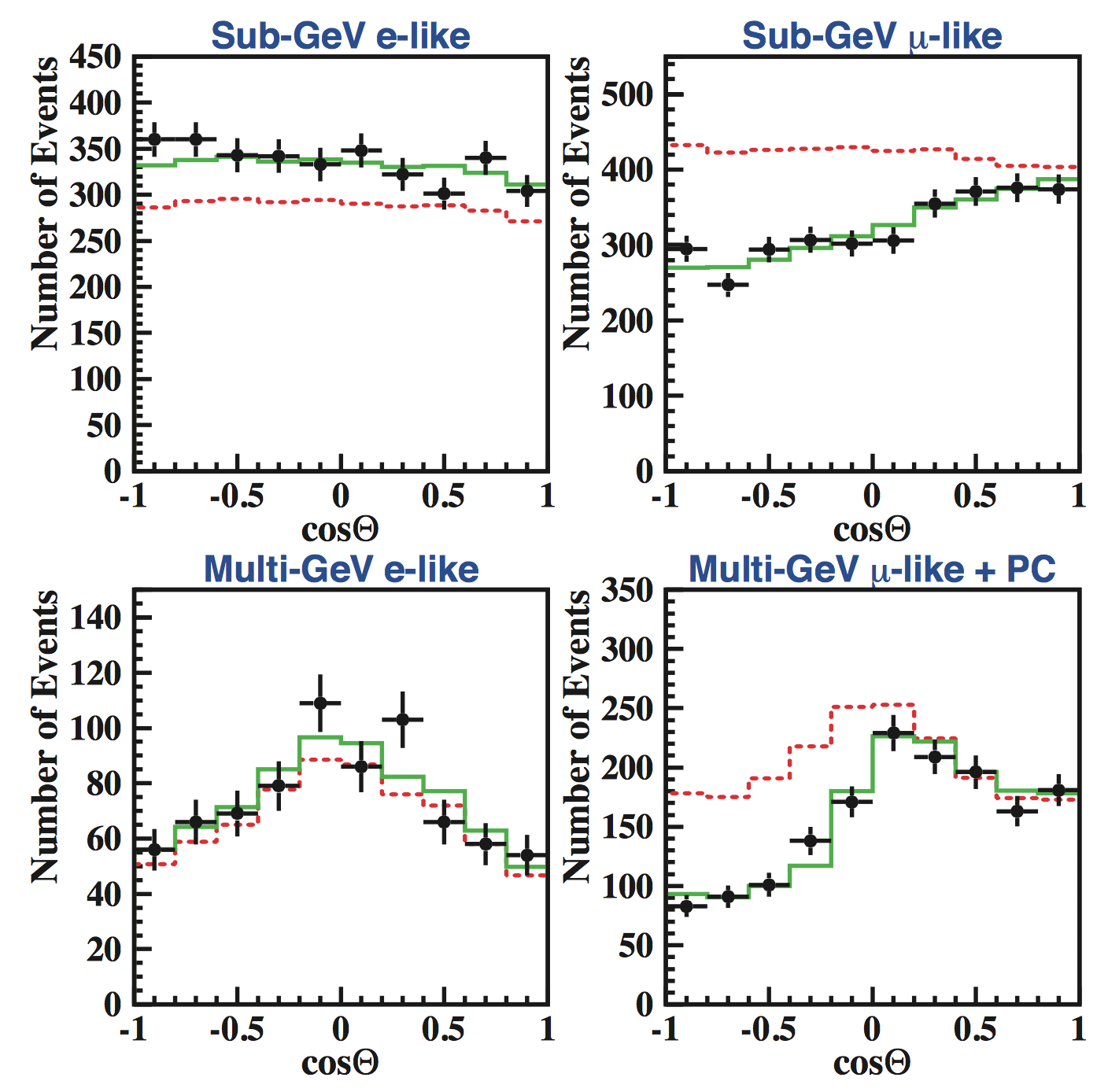}
\end{minipage}%
\caption{%
Zenith angle distributions for atmospheric neutrino events fully contained in
the SK detector, for electron- (left) and muon-like (right) events. The upper
(lower) plots are for visible energy below (above) 1.33\,GeV. The high-energy
muons are combined with partially contained events (PC). The error bars are the
measured data, the red dotted (green solid) lines show the prediction without
oscillations and with the best-fit oscillation scenario, respectively. Plot
taken from \cite{PDG2014}.}
\label{fig-sk-atm-nu}
\end{center}
\end{figure}

SK is installed in a cavern of a mine, with an overburden of 1000\,m of rock. 
The detector volume is a stainless-steel vessel, about 40\,m in diameter and
40\,m in height, filled with 50\,ktons of water. The outermost 18\,ktons are
used as veto layer against incoming charged particles. The inner volume is
observed by more than 11000 20-inch PMTs, the veto layer by about 1900 8-inch
PMTs. SK is operational since 1996, with a period of reduced sensitivity in
2001--2006 as the consequence of an accident destroying more than 50\% of the
large PMTs. SK has played a crucial role in the discovery and precision
investigation of neutrino oscillations. While one piece of evidence came from
confirming the solar neutrino deficit (i.e.\ the fact that less solar neutrinos
were measured than expected by the solar standard model), the breakthrough was
the observation of oscillations of atmospheric neutrinos (see
Fig.~\ref{fig-sk-atm-nu}) in 1998. The final confirmation that the solar
neutrino deficit is a neutrino flavour transition effect had to await the SNO
results (see below). See \cite{Takeuchi} for a summary of recent SK results.

SNO is located below an overburden of 2100\,m of rock in a deep mine in Ontario,
Canada. It uses a cavity with a diameter of 22\,m and a height of 34\,m. The
core of the detector is an acrylic vessel filled with 1\,kton of heavy water
(D$_2$O) and observed by more than 9000 8-inch PMTs. The remaining volume of the
cavern is filled with normal water (H$_2$O), serving as a veto volume against
incoming charged particles. The experiment started data taking in 1999 and was
operated until 2006. Currently a follow-up experimental phase (SNO+) is in
preparation, employing a liquid-scintillator filling doped with Tellurium for
the search for neutrino-less double beta decay.

\begin{figure}[ht]
\begin{center}
\begin{minipage}{\columnwidth}
  \includegraphics[width=\columnwidth]{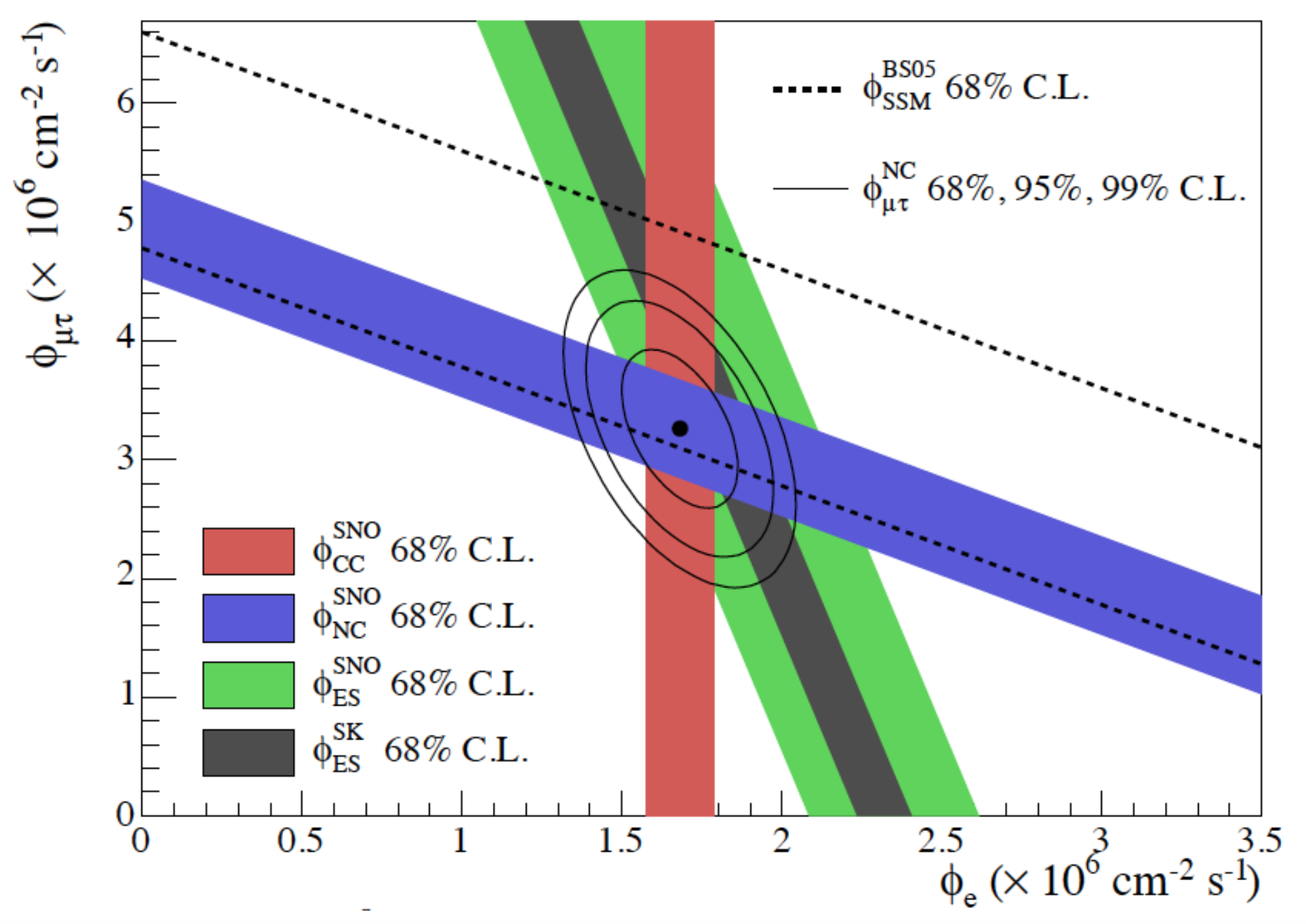}
\end{minipage}%
\caption{%
Solar $\nu_e$ vs.\ $\nu_\mu+\nu_\tau$ fluxes, with the experimental constraints
by SNO indicated by the coloured bands. The grey band shows the SK result.
Point and error contours are the combined result from the CC and NC measurements.
The dashed lines represent the prediction by the standard solar model.  Plot
taken from \cite{SNO2005}.}
\label{fig-sno-result}
\end{center}
\end{figure}

The deuteron target provides detectable reactions of all neutrino flavours:
$\nu_ed\to e^-pp$ (charged-current), $\nu_xd\to\nu_xpn$ (neutral-current) and
$\nu_xe^-\to\nu_xe^-$ (elastic, also measurable by SK). The secondary electrons
are observed via their Cherenkov light, the neutrons by gamma radiation emitted
when they are captured by deuterons or, in a later phase, by chlorine nuclei
added through a NaCl doping. The gamma radiation Compton-scatters on electrons
which then generate Cherenkov light. The measured rates of the three reactions
given above allow for a precise determination of both, the $\nu_e$ and the
$\nu_\mu+\nu_\tau$ fluxes from the Sun. The overall neutrino flux is found to be
consistent with the model expectation, but to consist only to about a third of
electron neutrinos (see Fig.~\ref{fig-sno-result}). This finding solved the
puzzle of the solar neutrino deficit and established a consistent standard
description of neutrino oscillations.

In 2015, the Nobel Prize in physics was awarded to Takaaki Kajita (SK) and
Arthur B.~McDonald (SNO) for {\it the discovery of neutrino oscillations, which
shows that neutrinos have mass}. Note that the Nobel Committee selected the
pictures reproduced in Figs.~\ref{fig-sk-atm-nu}~and~\ref{fig-sno-result} for
the corresponding announcement.
%
%
\subsection{Deep-ice and deep-water neutrino telescopes}
\label{sec-neu-tel}

For neutrino energies beyond some 100\,GeV, the detectors discussed in the
previous subsection lack sensitivity, simply because the target volume is too
small to yield sufficient event rates and also because events at such energies
are too large to be contained in the detector volume. In order to access this
high-energy regime, in particular for the purpose of neutrino astronomy,
instrumented volumes of at least several 10\,Mtons, better Gtons (i.e.\ cubic
kilometres of water/ice) are required. They are constructed by deploying arrays
of vertical structures (``strings'') carrying PMTs to the deep ice of the South
Pole or the deep water of the Mediterranean Sea or the Lake Baikal. The
water/ice layer above the sensors completely shields the daylight. The PMTs are
included in pressure-resistant glass spheres which also house the voltage
supplies, the front-end electronics and calibration instruments (optical
modules). The strings are connected to surface/shore by cables for data
transport, operation control and electrical power supply. Particle-induced
events from neutrinos or atmospheric muons are recognised and reconstructed
using the space-time pattern of Cherenkov photons recorded by the optical
modules. See \cite{KatzSpiering2012} for more details.

The main science objectives of these neutrino telescopes include neutrino
astronomy (i.e.\ observing the sky ``in the light of neutrinos'');
multi-messenger astronomy (i.e.\ combining the neutrino results with
electromagnetic and gravitational wave observations); the indirect search for
dark matter; neutrino and other particle physics (neutrino oscillations,
neutrino interactions, etc.), the search for phenomena beyond the standard model
of particle physics (magnetic monopoles, violation of Lorentz invariance,
sterile neutrinos, etc.). Note that the Universe is transparent to neutrinos of
all energies, whereas the reach of electromagnetic radiation is severely
constrained for energies exceeding some 10\,TeV due to gamma-ray interaction with
ubiquitous radiation fields.

\begin{figure}[t]
\begin{center}
\begin{minipage}{\columnwidth}
  \includegraphics[width=\columnwidth]{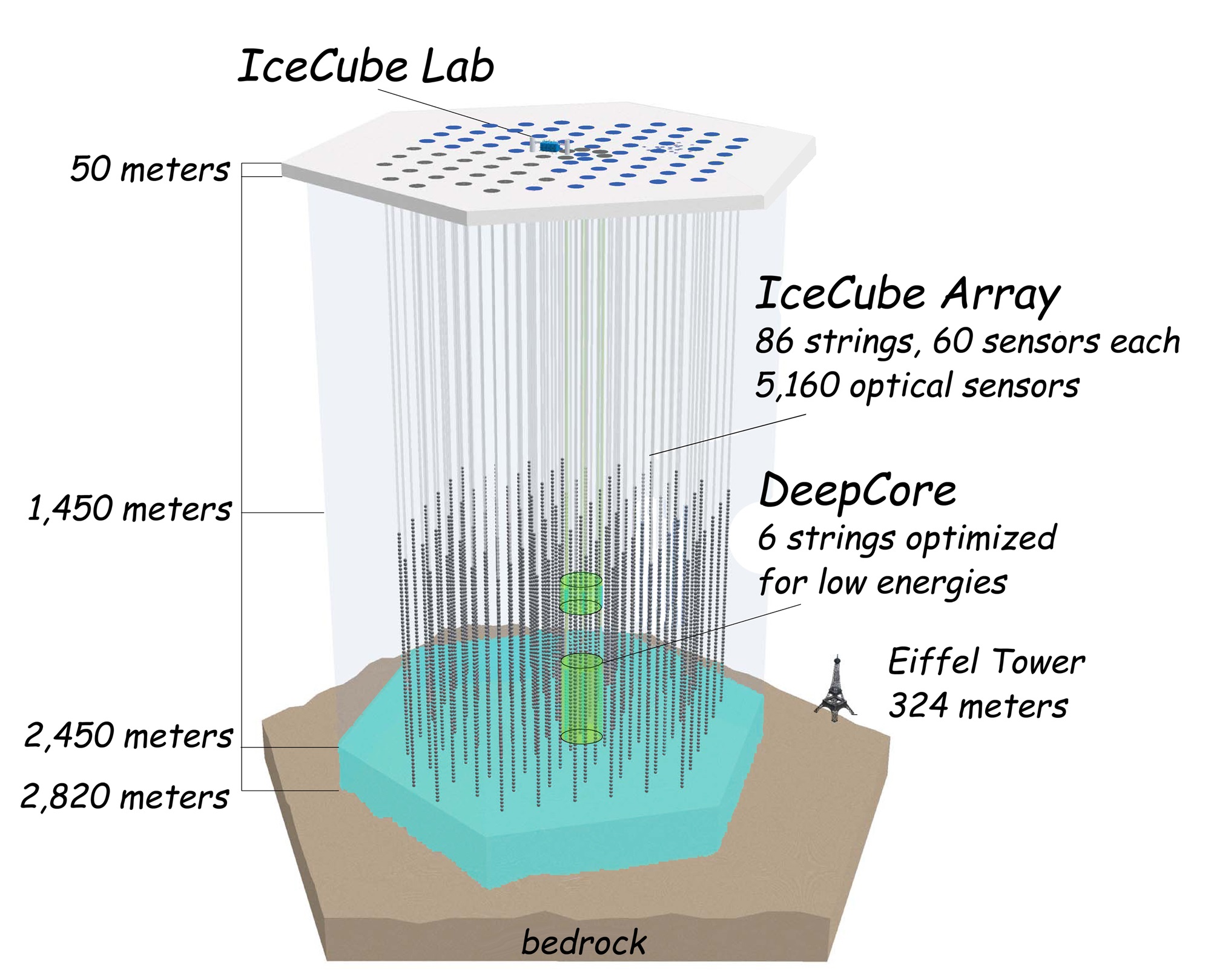}
\end{minipage}%
\caption{%
Schematic of the IceCube detector. Picture provided by the IceCube
Collaboration.}
\label{fig-icecube}
\end{center}
\end{figure}

The currently most sensitive high-energy neutrino telescope is IceCube
\cite{IceCube} at the South Pole, operational in full configuration since 2010. 
It consists of 86 strings carrying altogether 5160 downward-looking 10-inch PMTs
instrumenting one cubic kilometre of ice at a depth between 1450\,m and 2450\,m. 
A sub-volume is instrumented more densely than the rest to detect neutrinos with
energies down to 10\,GeV (Deep Core, fiducial volume about 4\,Mtons). The IceTop
Cherenkov surface array (frozen water tanks with optical modules) serves for
cosmic-ray studies and also provides some veto capability against muons and
neutrinos from air showers. An in-ice hardware trigger requires hits in adjacent
PMTs within $1\,\upmu s$. A schematic of IceCube is shown in
Fig.~\ref{fig-icecube}. The angular resolution for $\nu_\mu$ charged-current
events above some TeV is a few tenths of a degree; cascade events resulting from
reactions of other neutrino flavours or neutral-current interactions are
reconstructed with a resolution of about 10--15$^\circ$. Main systematics come
from the inhomogeneity of the optical ice properties and from light scattering,
which blurs the space-time pattern of Cherenkov photons.

IceCube has achieved two major breakthrough results: The first identification
and flux measurement of high-energy cosmic neutrinos (2013) and the first
association of high-energy neutrinos to an astrophysical object, the blazar
TXS\,0506.056 (2018). This detection became possible by relating an IceCube
neutrino alert with electromagnetic observations, and was confirmed by archival
IceCube data. See \cite{Williams,Williams2018} for these and further IceCube
results.

The IceCube detector will be further developed and extended. As a first step,
7~additional strings with newly developed optical modules and calibration
devices will be added to the Deep Core region. This project has currently been
approved by the US National Science Foundation (NSF), and deployment is expected
in 2022/23. The main objectives are a better understanding of the ice properties,
entailing a reduction of the systematic uncertainties; improved investigation of
neutrinos in the few-GeV range; test of new hardware developments. As a next
step, IceCube-Gen2 with a 10\,km$^3$ deep-ice detector, a high-density core for
low-energy neutrinos (PINGU), a large cosmic-ray and veto surface array, and a
radio detection array is planned. If all works well, IceCube-Gen2 could be
installed 2025--2031. See \cite{Williams} for details.

Complementing IceCube in the field of view and in the major systematic
uncertainties, the ANTARES neutrino telescope in the Mediterranean Sea off the
French shore near Toulon is operational in full configuration since 2008. It
consists of 12 strings carrying 25 storeys with three 10-inch PMTs each,
downward-looking at an angle if $45^\circ$ to vertical. The strings are
connected to a junction box on the sea bed and from there by an electro-optical
cable to shore. All PMT hits exceeding a signal height corresponding to
0.3~photo-electrons are read out and sent to shore, where the data are filtered
by an online computer cluster. ANTARES instruments a water volume of about
0.015\,km$^3$ and is thus intrinsically significantly less sensitive than
IceCube. Angular resolutions for $\nu_\mu$ charged-current and for cascade
events with neutrino energies exceeding 10\,TeV are better than $0.4^\circ$ and
2--3$^\circ$, respectively. Main instrumental systematics are due to the
inhomogeneity of detector and deep-sea environment in time, and due to
background light from bioluminescence. In spite of its limited size, ANTARES
has provided a number of important results, in particular also in common
analyses with IceCube, which evidence the importance of full sky coverage for
each neutrino flavour and energy. See \cite{Chiarusi} for a selection of
important ANTARES results.

ANTARES has proven the feasibility of deep-sea neutrino detection and has paved
the way towards the next-generation neutrino telescope in the Mediterranean Sea,
KM3NeT \cite{KM3NeT-LoI2016}. The KM3NeT~2.0 detector will consist of two
installations, ARCA\footnote{Astroparticle research with cosmics in the abyss}
off the eastern Sicilian shore and ORCA\footnote{Oscillation research with
cosmics in the abyss} close to the ANTARES site. ARCA will encompass two
building blocks with 115~strings each, where each string carries 18~optical
modules; the overall instrumented volume will be 1\,km$^3$. The prime objective
of ARCA is neutrino astronomy in an energy range beyond a few TeV. Due to its
larger size and improved detector technology (see below), the angular
resolutions in KM3NeT will be better than in ANTARES ($<0.1^\circ$ for $\nu_\mu$
and $<2^\circ$ for cascade events at energies of about 100\,TeV and above). ORCA
will use the same basic detector technology as ARCA, but be much more densely
instrumented (115 strings, 18 optical modules per string, instrumented volume
0.06\,km$^3$). ORCA \cite{Brunner2018} will focus on neutrino oscillation
physics with atmospheric neutrinos in the energy range of a few GeV to a few
10\,GeV, and in particular on measuring the neutrino mass ordering. An option to
investigate CP violation by directing a neutrino beam from Protvino to ORCA
(P2O) \cite{Zaborov2018} is under discussion.

\begin{figure}[t]
\begin{center}
\begin{minipage}{0.6\columnwidth}
  \includegraphics[width=\columnwidth]{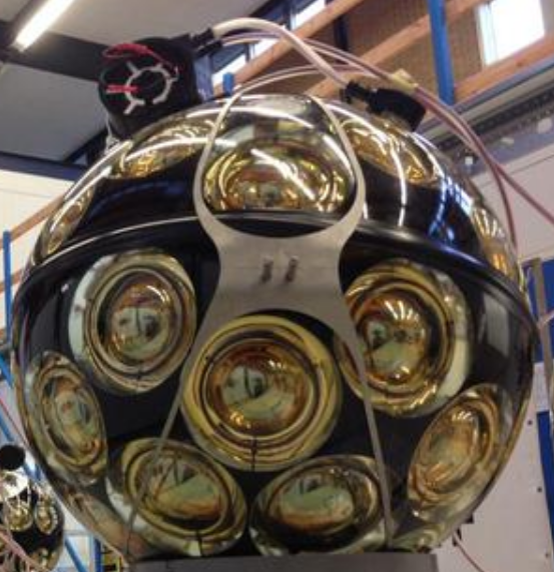}
\end{minipage}%
\caption{%
Photograph of a KM3NeT digital optical module during assembly/testing. 
Picture provided by the KM3NeT Collaboration.}
\label{fig-km3net-dom}
\end{center}
\end{figure}

A number of new technical developments has been achieved for KM3NeT that improve
functionality, cost-effectiveness, risk mitigation and construction time. 
Amongst them are equi-pressure vertical cables (electrical leads and optical
fibres in an oil-filled hose), a new deployment strategy (strings wound on a
spherical structure which is deployed to the sea floor and then unfurls upon
release), and a multi-PMT digital optical module (see
Fig.~\ref{fig-km3net-dom}). The advantages of the latter are increased
photo-cathode area per glass sphere at reduced overall PMT cost; reduced risk
due to less feed-through holes in the glass spheres; better photon-counting;
angular information; almost isotropic sensitivity. The construction of both
KM3NeT detectors has begun and is expected to be completed 2021/22. A further
extension with four more ARCA blocks is envisioned but not yet negotiated. See
\cite{Chiarusi} for expected KM3NeT sensitivities, in particular to a diffuse
cosmic neutrino flux as observed by IceCube and to the neutrino mass ordering.

A third large neutrino telescope, the Gigaton Volume Detector (GVD), is
currently under construction in Lake Baikal, Russia. It will consist of 8-string
clusters with a diameter of 120\,m and a height of 525\,m; the depth at the
installation site is 1360\,m. Each string carries 36~optical modules equipped
with 10-inch, downward-looking PMTs. In a first phase, 8 clusters will be
deployed to instrument 0.4\,km$^3$ of water; the final goal are 27~clusters
covering a volume of 1.5\,km$^3$. Currently three clusters are operational, and
two more are to be deployed per year. First results from GVD have been reported
at the Neutrino Conference 2018 \cite{Dzhilkibaev2018}.

%
%
\subsection{Other neutrino detectors}
\label{sec-neu-oth}

A new approach to detecting highest-energy $\nu_\tau$ is the use of IACTs
(see~Sect.~\ref{sec-gam}) directed on the sea surface or other surface areas of
the Earth. The basic principle is that $\nu_\tau$ enter the Earth under zenith
angles close to 90$^\circ$, undergo a charged-current reaction and produce a
$\tau^\pm$ that enters the atmosphere and initiates an extended air shower upon
its decay. A first study of this option has been performed with MAGIC
\cite{Mirzoyan,MAGIC2018}; other experimental approaches to detect tau neutrinos
through the same mechanism are discussed in \cite{Alvarez-Muniz2018}.

%
%
\section{Cosmic-ray and hybrid detectors}
\label{sec-cos}

In several large detector infrastructures, Cherenkov detection is combined with
other measurement methods, such as the observation of fluorescence light, direct
particle detection with scintillator or other instruments, or radio detection. 
Corresponding hybrid infrastructures are typically targeting cosmic rays, in
several cases in combination with gamma-ray and/or neutrino measurements.

\begin{figure}[ht]
\begin{center}
\begin{minipage}{\columnwidth}
  \centering
  \includegraphics[width=0.9\columnwidth]{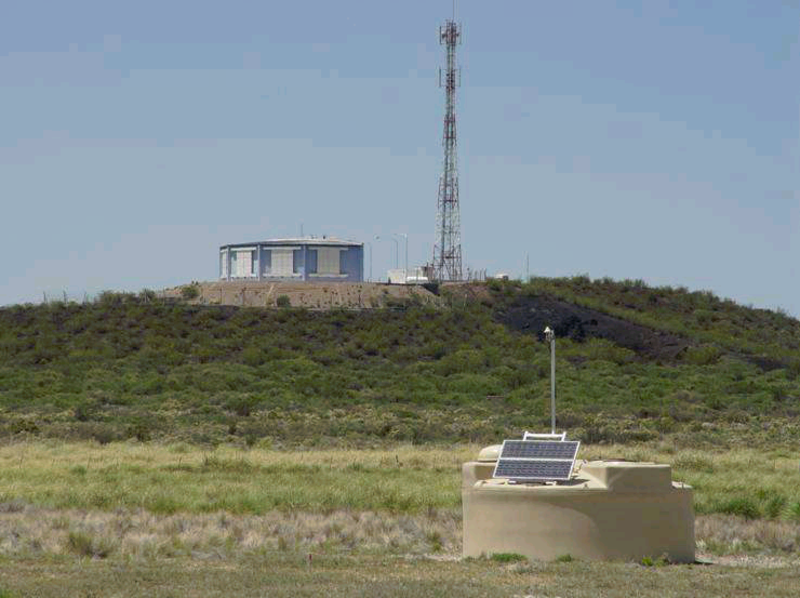}
\end{minipage}%
\caption{%
Photograph of a water Cherenkov detector (foreground) and a building housing six
fluorescence telescopes (background) of the Auger infrastructure. Picture provided
by the Auger Collaboration.}
\label{fig-auger-photo}
\end{center}
\end{figure}

The largest hybrid detector system to date is the Pierre Auger Observatory
(``Auger'') \cite{Auger} in Argentina, combining 1660 PMT-equipped water tanks
on a 3000\,km$^2$ area for the direct detection of charged particle from air
showers with 27 fluorescence telescopes observing the atmosphere above the
ground array (see Fig.~\ref{fig-auger-photo}). Auger targets cosmic rays with
energies of about $10^{17.5}$--$10^{21}$\,eV and searches for gamma-rays and
neutrinos in the same energy interval. The water detectors yield a measurement
of the ``footprint'' of an air shower, from which the energy can be inferred,
and act as timing arrays (see Sect.~\ref{sec-gam-tim}) for the direction
measurement. They have a duty cycle of close to 100\%. The fluorescence
telescopes measure the light intensity along the shower and, in stereoscopic
observations, determine shower position and direction. They thus provide an
independent determination of energy and direction, which is cross-calibrated
against and combined with the array measurement. Even though the fluorescence
telescopes can only be operated in clear, moon-less nights and thus have a
limited duty cycle, their contribution to the control of systematics and thus to
the resulting experimental precision is essential.

\begin{figure}[ht]
\begin{center}
\begin{minipage}{\columnwidth}
  \includegraphics[width=\columnwidth]{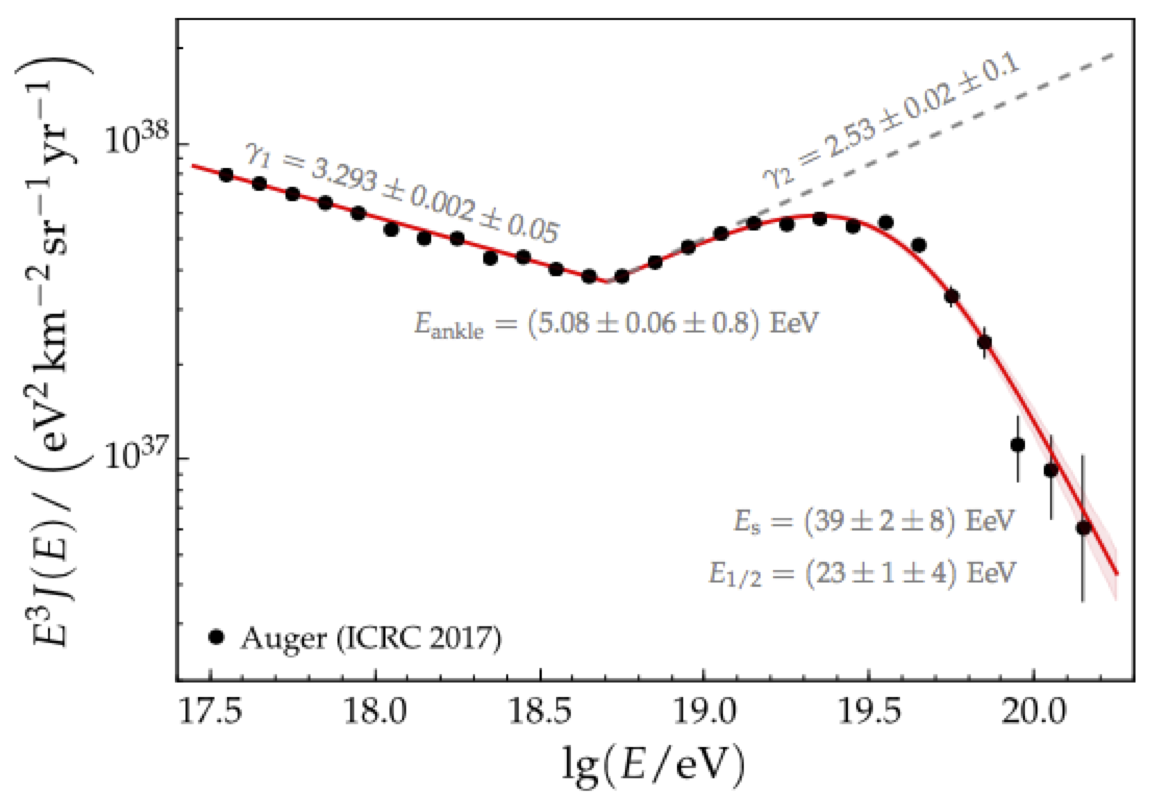}
\end{minipage}%
\caption{%
Cosmic-ray energy spectrum beyond $10^{17.5}\,$eV measured by Auger. The power-law
behaviour below and above the ``ankle'' is indicated by the dashed lines. Beyond
$10^{19.2}\,$eV the spectrum steeply decreases. Picture from
\cite{AugerICRC2017}.}
\label{fig-auger-spectrum}
\end{center}
\end{figure}

Auger started operation with a partial installation in 2001 and has achieved major
breakthroughs, amongst others a precise measurement of the cosmic ray spectrum
above $10^{17.5}$\,eV indicating a clear cut-off beyond $10^{19.2}$\,eV (see
Fig.~\ref{fig-auger-spectrum}), and the first detection of a dipole anisotropy
of the arrival direction of cosmic rays at highest energies
($E>8\times10^{18}\,$eV) \cite{Auger2018}. See \cite{Maris} for a summary of
these and further recent results. Still, important questions remain open,
amongst them the chemical composition of cosmic rays, the origin of said
anisotropy, and the detection of cosmogenic gamma-rays from interactions of
highest-energy protons/nuclei with the cosmic microwave background. Auger is
currently upgraded to ``AugerPrime'' with new fast electronics and in particular
with scintillator detectors on top of the Cherenkov tanks to improve the
discrimination between hadronic and leptonic shower components on ground.

The Tunka experiment in the Tunka valley near Lake Baikal and its current
extension stage TAIGA (Tunka Advanced Instrument for cosmic ray physics and
Gamma Astronomy) \cite{TAIGA,Budnev2018} comprise five different detector
systems: An array of 133 PMTs observing the sky at clear moon-less nights
(Tunka-133), a radio array (Tunka-Rex), an array of 18 scintillation stations
recording charged particles (Tunka-Grande, see also \cite{Vaidyanathan}),
55~wide-angle Cherenkov stations providing improved Cherenkov light detection
(TAIGA-HiSCORE) and several IACTs under construction (TAIGA-IACT).

A different type of hybrid detector is NEVOD \cite{NEVOD}, a comparatively small
ground-level installation in Moscow, Russia. It is unique in combining a
2.1\,kton water volume observed by photomultipliers in 91~multi-PMT optical
modules with additional scintillator detector tiles for calibration and streamer
tube tracking planes for precisely measuring the trajectories of muons or muon
bundles. The combination of muon tracking and Cherenkov measurements offers
promising options to cross-check simulations and calibrate e.g.\ acceptances of
optical modules. Also, the simultaneous measurement of muon directions and
energies (via the Cherenkov intensity) can be interesting for cosmic-ray
studies. See \cite{Petrukhin} for more details.

Future hybrid approaches are the LHAASO project in China, targeting gamma-rays
\cite{Vernetto2016}, and the space missions JEM-EUSO \cite{JEM-EUSO} and
POEMMA \cite{Olinto2018}, both targeting ultra-high energy cosmic rays and
neutrinos.

%
%
\section{Conclusion and Outlook}
\label{sec-con}

Cherenkov detectors play a crucial role in gamma-ray, neutrino and cosmic-ray
astroparticle physics. Many of the recent breakthrough-results would not have
been possible without them, and they are essential for the future experiments
being constructed or planned.

%
%
\raggedright
\bibliography{rich2018_katz}

\begin{thebibliography}{10}
\bibitem{Petrukhin}
{A.~Petrukhin}, Cherenkov water detector {NEVOD} and its further development,
  in: this issue.
\end{thebibliography}
\end{document}